\documentclass[useAMS,usenatbib,twocolumn]{mnras}

\usepackage[]{color,graphicx}
 
\usepackage{amsmath}

\def\lesssim{\mathrel{\hbox{\rlap{\hbox{\lower5pt\hbox{$\sim$}}}\hbox{$<$}}}}
\def\gtrsim{\mathrel{\hbox{\rlap{\hbox{\lower5pt\hbox{$\sim$}}}\hbox{$>$}}}}

\title[Supernova forecast with strong lensing]
{Supernova forecast with strong lensing}
\author[Y. Suwa]{
Yudai Suwa$^{1}$\thanks{E-mail: suwa@yukawa.kyoto-u.ac.jp}
\\
$^{1}$Center for Gravitational Physics, Yukawa Institute for Theoretical Physics, Kyoto University, Kyoto, 606-8502, Japan
}

\begin{document}

\date{Accepted. Received.}

\pagerange{\pageref{firstpage}--\pageref{lastpage}} \pubyear{2017}

\maketitle

\label{firstpage}

\begin{abstract}

In the coming LSST era, we will observe $\mathcal{O}(100)$ of lensed
supernovae (SNe). In this paper, we investigate possibility for
predicting time and sky position of a supernova using strong
lensing. We find that it will be possible to predict the time and
position of the fourth image of SNe which produce four images by
strong lensing, with combined information from the three previous
images. It is useful to perform multi-messenger observations of the
very early phase of supernova explosions including the shock breakout.

\end{abstract}


\section{Introduction}
\label{sec:intro}

Supernovae (SNe) represent the final phase of a massive star. When a
shock that forms inside the star pierces the stellar surface, so
called {\it shock breakout} (SBO) takes place, which makes a bright
flush in X-ray/ultraviolet bands \citep[see][for a recent
  review]{waxm16}. SBO and the early light curve (LC) of SN are useful
probes to investigate the very final stage of stellar evolution. There
have been three (including candidate) observations of SBO so far
(SN2008D, \citealt{sode08b}; SNLS-04D2dc, \citealt{scha08}; KSN 2011d,
\citealt{garn16}). SN2008D was serendipitously observed by Swift, and
others were found by survey programs.

The biggest survey volume in optical/infrared bands is achieved by
Subaru Hyper Suprime-Cam \citep{miya12} at the current moment and will
be made by the Large Synoptic Survey Telescope (LSST) \citep{mars17}
in the near future. There are also other survey programs, such as the
Palomar Transient Factory (PTF, \citealt{law09}), Panoramic Survey
Telescope and Rapid Response System (Pan-STARRS1, \citealt{kais10}),
the High Cadence Transient Survey (HITS, \citealt{foer16}), the Kiso
Supernova Survey (KISS, \citealt{moro14}), etc. These surveys will
eventually detect SBO soon. Follow-up observations are possible only
after the alert sent by these survey programs, so that multi-telescope
observations of SBO is very difficult even with these magnificent
survey programs. Accordingly, we propose a different way to catch SBO
using strong lensing.

After more than 50 years since the original idea by \citet{refs64},
there have been three observations of strongly lensed SNe. The first
one was PS1-10afx, which is a Type Ia SN. This event was not detected
as resolved multiple images, but it was magnified by a factor of 30
because of strong lensing \citep{quim13,quim14}. The next one was SN
Refsdal, which is an SN 1987-like core collapse SN. This was the first
ever lensed SN with {\it resolved} multiple images
\citep{kell15}. Importantly, from the positions of these images, it
was predicted that there would be an additional image one year {\it
  after} the discovery of them and it indeed appeared
\citep{kell16}. This is the first {\it SN forecast}. The third one was
iPTF16geu, which is a Type Ia SN. At first, multiple images were not
resolved by PTF, but follow-up observations by Hubble Space Telescope
(HST) resolved multiple images, so that this is the first lensed Type
Ia SN with resolved multiple images \citep{goob17}. It clearly
demonstrated the usefulness of observations of multi-telescope for
lensed SNe.

Here, we mention SN Refsdal a little bit more in detail. This is an SN
occurring at $z=1.49$ and lensed by galaxy cluster at $z=0.54$. The SN
itself was lensed by a galaxy between host galaxy and observer, and
found as four images around host galaxy. Furthermore, the SN host
galaxy was also lensed by the cluster, producing three
images. Follow-up analysis suggested that {\it two} more images were
possible, in addition to the already observed four images
\citep[e.g.][]{ogur15,treu16}. One of them would have appeared 17
years ago, while another one would appear one year after the original
four images. This case allowed a stringent test of strong lensing
models. Indeed, a new image was found at the predicted time and sky
location.

It was predicted that LSST will observe more than 100 {\it strongly}
lensed SNe \citep{ogur10}. This is actually a conservative estimate
and the number can be even more than 1000, depending on the criteria
(M. Oguri, private communication). In this work, we will demonstrate
the viability of SN forecast and multi-messenger observations of SBO,
using lensed SNe.

\section{Time delay, magnitude, and separation}
\label{sec:time_delay}

In this work, we specify LSST for triggering new SNe. LSST is a survey
program, which uses a telescope with 8.4 m mirror, 15 s exposure, 9.6
deg$^2$ of field of view, and six bands. It sweeps whole sky with
$\approx 24$ mag of $5\sigma$ depth, a typical seeing full width at
half-maximum (FWHM) being 0.75 arcsec, and 5 day cadence. A number of
transient astrophysical objects will be detected, for instance,
$\mathcal{O}(10^6)$ of SNe, during its ten-year survey.  A predicted
number of lensed SN, which will be observed by LSST is $\sim$130
\citep{ogur10}. Among the lensed SNe, the fraction of Type Ia is
$\sim$ 34\%,\footnote{The number is slightly different from value in
  Table 3 of \citet{ogur10}, because of Poisson noise of sampling.}
Ib/c is $\sim$ 31\%, IIL is $\sim$5\%, IIP is $\sim$ 15\%, and IIn is
$\sim$ 15\%.  In this work we focus on Type Ib/c SNe, since they have
a relatively short timescale in LC evolution because of their small
ejecta mass, and hence they make the systematic error of the SBO time
estimation determined from the LC smaller than for other SN types.

\cite{ogur10} conducted Monte-Carlo calculations regarding quasar and
SN lensing, and estimated numbers of predicted lensed transient
objects for various survey programs. Since they made their mock data
public,\footnote{Since it was not available online when the author
  performed this work, we asked M. Oguri to provide it.} we use their
data in this work. In their calculations, they took into account not
only limiting magnitudes, but also angular resolutions to resolve
multiple images. They omitted images with separation being smaller
than 2/3 of the seeing FWHM of the survey programs (seeing FWHM being
0.75 arcsec for LSST).  After their prediction, iPTF16geu was observed
without resolving multiple images by the initial survey telescope
(PTF), but it was resolved into multiple images by a follow-up
observation done by HST. Therefore, the number of lensed objects
resolved into multiple images can be significantly larger than their
original estimate, i.e. their estimate is rather conservative.

In the following, we investigate lensed Type Ib/c SNe in mock catalog
by \cite{ogur10}.  Their mock sample limits events whose third
brightest images being brighter than 22.6 mag, which is 0.7 mag
brighter than the magnitude limit of each visit (23.3 mag for LSST).
Table \ref{tab} summaries the properties.  Among their 1219 lensed SN
samples, which are ten times oversampled, we select 376 Type Ib/c
SNe. 86 of them have four images, so that in the LSST era we will
observe roughly one SN with four images every year. It should be noted
that the mean peak magnitude $\left<m_\mathrm{lens}\right>$ does not
depend on the number of images.

\begin{table}
\caption{Summary of mock Type Ib/c SN data from \citet{ogur10}. The
  first half gives characteristics of all events, and the second half
  gives events having four images. $\left<\cdots\right>$ gives mean
  value with the standard deviation. $z_s$ is source redshift, $z_l$
  is lens redshift, $m_\mathrm{org}$ is original peak apparent
  magnitude without lensing, $m_\mathrm{lens}$ is peak apparent
  magnitude with lensing, $\theta_\mathrm{lens}$ is separation from
  the first image, and $t_\mathrm{delay}$ is the time delay of the
  fourth image from the first image.}  
\centering
\begin{tabular}{ll}
\hline
 all lensed Type Ib/c SNe\\
 \hline
 number of SNe &         376 \\
$\left<z_s\right>$ & 0.789 $\pm$ 0.231\\
$\left<z_l\right>$ & 0.326 $\pm$ 0.166\\
$\left<m_\mathrm{org}\right>$ &  22.6 $\pm$  1.02\\
 number of images &         926 \\
$\left<m_\mathrm{lens}\right>$ &  21.6 $\pm$  1.00\\
 \hline
 lensed Type Ib/c SNe with 4 images\\
 \hline
 number of SNe &          86 \\
$\left<m_\mathrm{lens}\right>$ &  21.7 $\pm$  1.18\\
$\left<\theta_\mathrm{lens}\right>$ (arcsec)$^a$ &  1.24 $\pm$  0.61\\
$\left<\mathrm{log}_{10}(t_\mathrm{delay}/\mathrm{day})\right>$$^b$ &  1.02 $\pm$  0.49\\
\hline
\end{tabular}
\label{tab}
\begin{flushleft}
$^a$ Mean separation from the first image, for second, third and fourth images.\\
$^b$ Mean logarithmic time delay between the first and fourth images.
\end{flushleft}
\end{table}%

\begin{figure}
\centering
\includegraphics[width=0.4\textwidth]{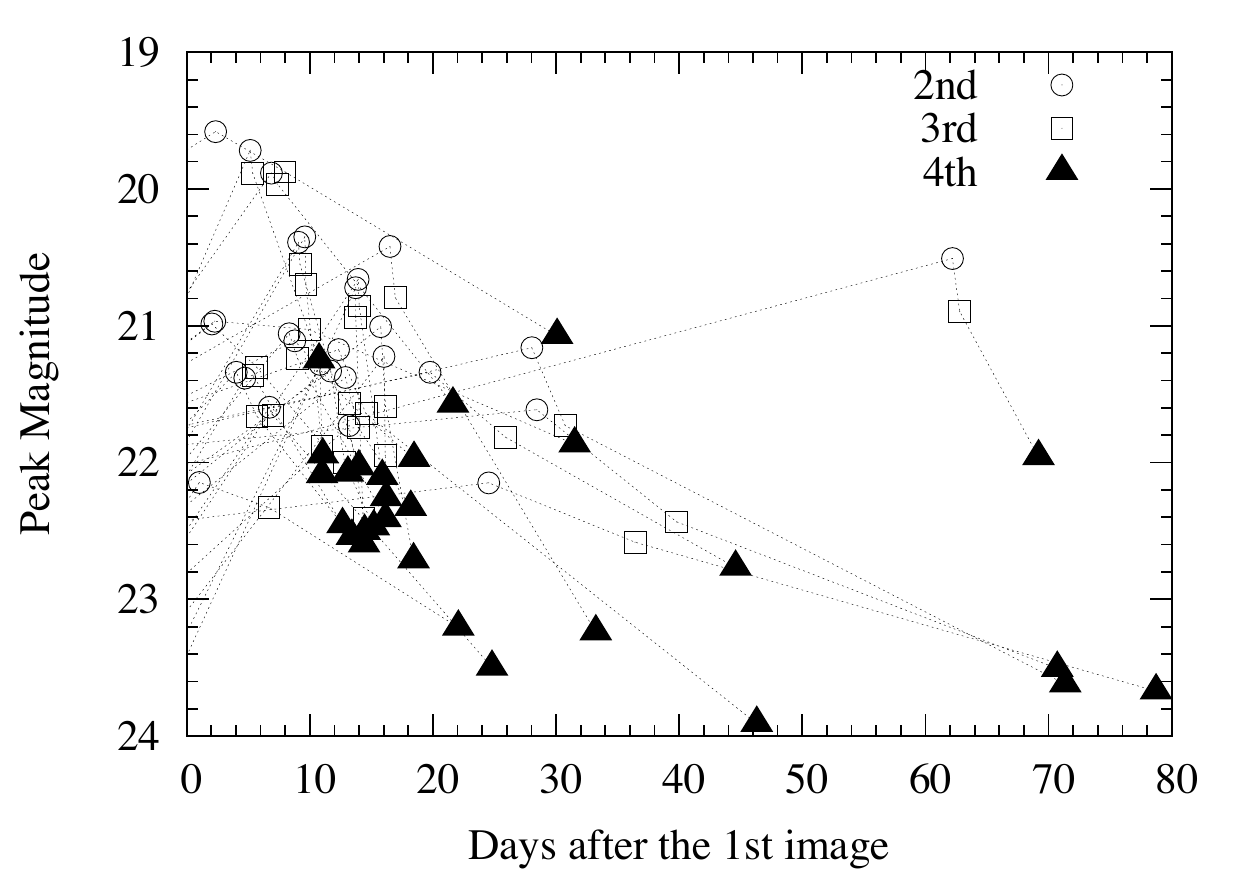}
\includegraphics[width=0.4\textwidth]{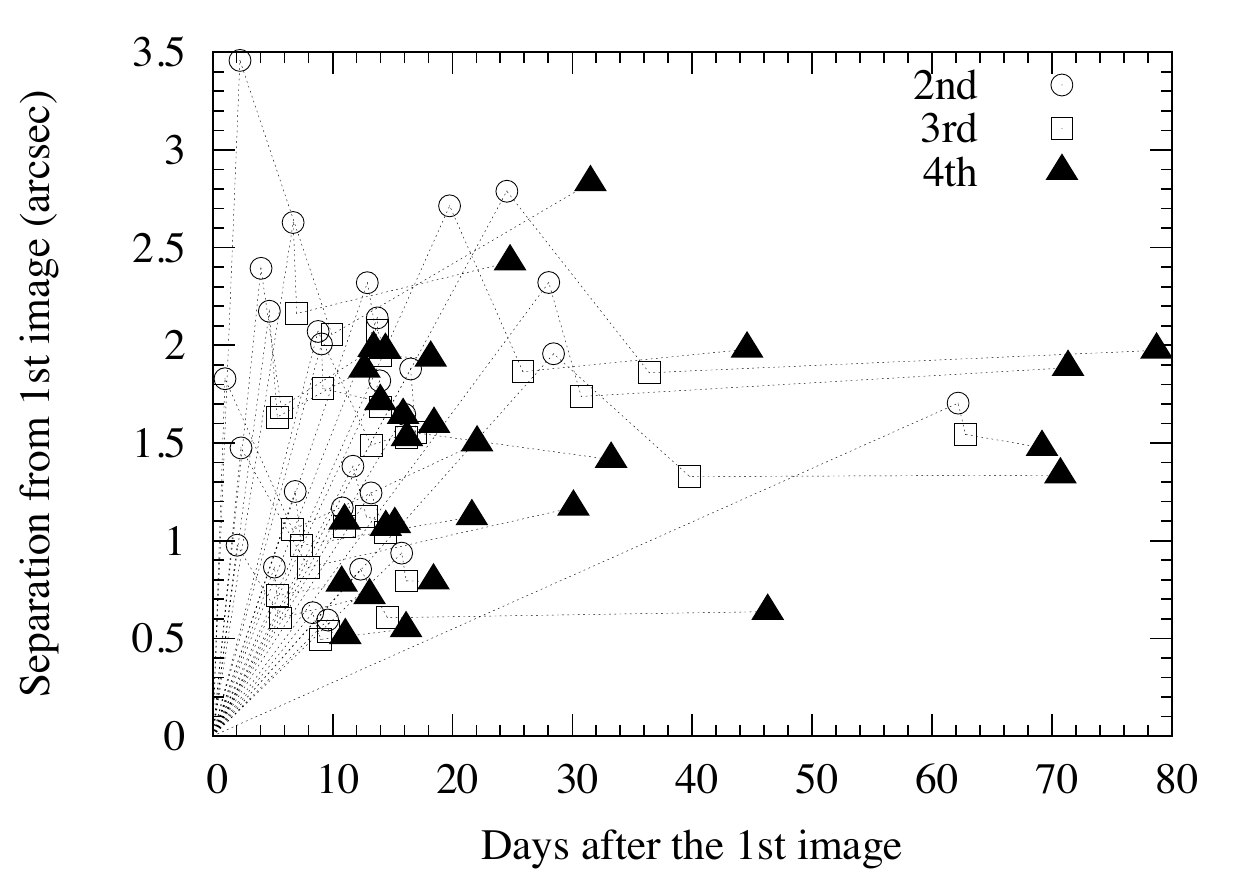}
\caption{ (Top) Peak magnitude of each image as a function of time
  delay from the first image. Among Type Ib/c SNe given by
  \citet{ogur10}, 28 events, which produce third images later than
  five days and fourth images later than ten days but no later than a
  hundred days after the first image, are shown. Open circles, open
  squares, and closed triangles indicate second, third and fourth
  images, respectively. Dashed lines connect images from the same SNe.
  (Bottom) The same as top panel, but for separation from the first
  image.}
\label{fig:t-mag}
\end{figure}

The top panel of Figure \ref{fig:t-mag} presents the peak magnitudes
of the second, third, and fourth images as a function of delay time
from the first image, for lensed SNe with four images. As is seen from
the figure, longer delay time leads to a dimmer image.  In some
systems, these fourth images appear $\mathcal{O}(10)$ days after the
first image emerged with magnitude brighter than 24, which can be
observed by 2--4 m-size telescopes. The bottom panel of the figure
indicates the angular separation of images from the first image. It
shows a large scatter and no apparent correlation with time
delay. Typical separations are about 1 arcsec, which can be resolved
by ground-based telescopes.

\begin{figure}
\centering
\includegraphics[width=0.45\textwidth]{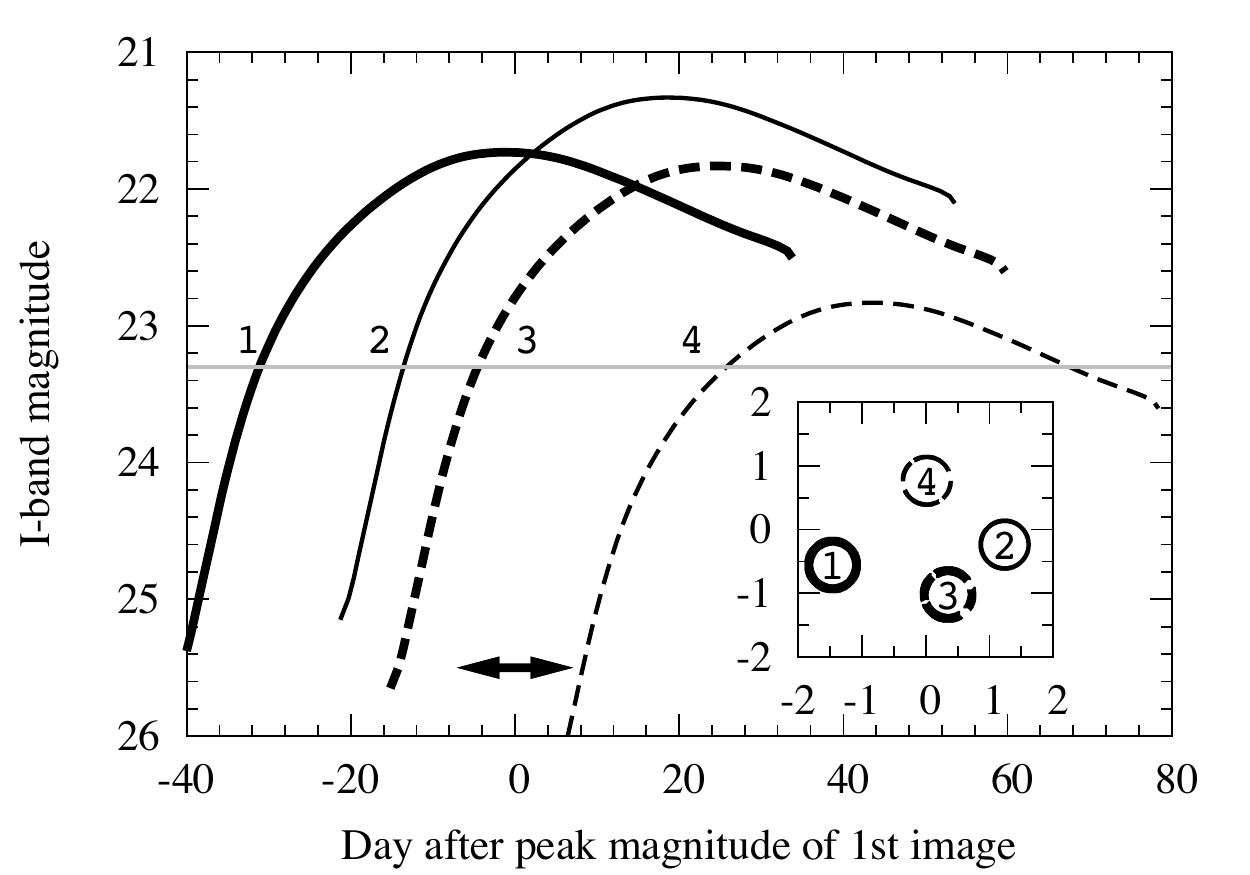}
\caption{An example light curve in i-band of an event with $z_s=0.87$
  and $z_l=0.314$. Light curve shape is taken from a typical Type Ib/c
  SN, SN1999ex. First, second, third, and fourth images correspond to
  thick-solid, thin-solid, thick-dashed, and thin-dashed lines,
  respectively. Time delays from the first image are 19.8 (second),
  25.9 (third), and 44.6 (fourth) days.  The arrow indicates the time
  of SBO of the fourth image, which is bright in UV/X-ray bands for
  the case of a Type Ib/c SN. Horizontal grey line indicates limiting
  magnitude of LSST.
  In the small panel, spacial positions of images are shown with a
  typical seeing FWHM of LSST (0.75 arcsec). Numbers in circles
  present corresponding image in LC.  }
\label{fig:sn1999ex}
\end{figure}

Figure \ref{fig:sn1999ex} presents an example LC of a lensed Type Ib/c
SN. The lens galaxy ID in the mock data is 7738038, which has
$z_s=0.87$ and $z_l=0.314$. The apparent peak magnitude in I-band
without lensing is 22.5 mag. Since \cite{ogur10} only took into
account the peak magnitude, we construct the LC by using a typical
Type Ib/c SN, SN1999ex \citep{stri02}. In the figure, the limiting
magnitude of LSST per visit (10$\sigma$) is shown as a grey horizontal
line. One can see that the first image can be observed about 30 days
before its peak and the second image will appear above the detection
threshold 18 days after the first image emergence. The third and
fourth ones will be found about 28 and 58 days after the first
image. In this case, the SBO emission of the fourth image will be
observed after the emergence of the third image.

\section{Strategy}
\label{sec:strategy}

In this section, we discuss a strategy to perform a multi-messenger
observation campaign for SBO of fourth images from lensed SNe. An
ideal scenario is as follows:
\begin{enumerate}
\item Find a new SN by LSST survey.
\item A second image appears $\mathcal{O}(1)$ d after the first image.
\item One calculates the lens potential based on these two images and
  predict the position and time of the third image.
\item By observing the third image with deeper and more frequent
  observations, one calibrates the lens potential model and LC
  evolution, and predicts the fourth image more precisely than the
  third one.
\item One targets the SBO of the fourth image using multiple
  telescopes.
\end{enumerate}

\begin{figure}
\centering
\includegraphics[width=0.3\textwidth]{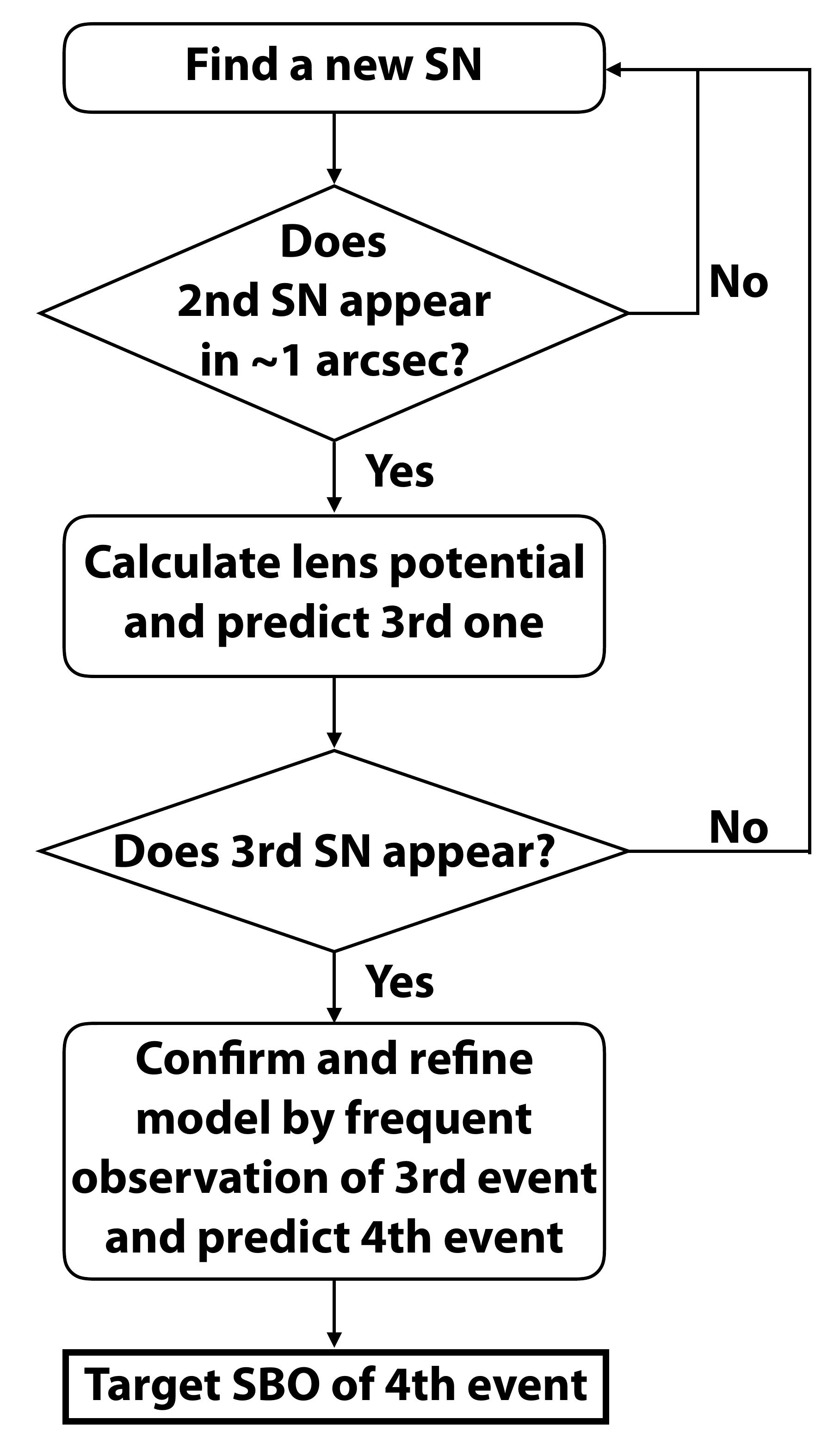}
\caption{A flowchart of the event selection. First and second images
  are found by LSST and third and fourth images are observed by other
  telescopes.}
\label{fig:schematic}
\end{figure}

In Figure \ref{fig:schematic}, we show our flowchart of the event
selection. In this figure, the first and second images should be
detected by LSST, and other telescopes are able to conduct more
frequent observations for third and fourth images.  In order to
observe the SBO, the most important part is the time precision of the
fourth image. If we can reduce the error, by intensively observing the
third image, up to $\lesssim 1$ d, the feasibility of the SBO
observation becomes remarkably high. For predicting the third image
properties, we need to determine at least five lens model parameters
if we employ the singular isothermal ellipsoid (SIE), which is used
most frequently to model lensing galaxies, assuming redshifts are
determined well by photometric data: velocity dispersion; ellipticity;
orientation of the lens galaxy; source position. Up to six degrees of
freedom can be fixed by observation of the first two images: two image
positions; flux ratio; time delay. Hence, in principle, the prediction
is possible.  Of course, for a more realistic lens model, there are
more parameters to be determined. But on the other hand, the
observation of the third image does provide significantly more
information since, together with the first and second image, there are
now twelve observables: three image positions; three flux ratios;
three time delays.
It should also be noted that we are neglecting host galaxy distortion,
which would provide additional information, so that our results in the
following are conservative.

By using the same mock data as Figure \ref{fig:sn1999ex}, we attempt
to {\it predict} the time delay and position of the fourth image from
information of the previous three images with {\tt glafic}
\citep{ogur10b}, which is a public software package for analyzing
gravitational lensing (see Appendix \ref{sec:glafic}). A similar study
was done in \citet{ogur03}, but it was more interested in cosmological
applications. As input data, we employ sky positions of three images,
the redshift of host galaxy, flux ratios, and time delays with respect
to the first image, with 1$\sigma$ errors of 0.75 arcsec, 0.5, 50\%,
and 5 days, respectively. The best fit is obtained with the delay time
for the fourth image being 44.74 days, which is slightly (0.17 day)
later than the {\it correct} value.\footnote{The same analysis for the
  third image using information of the first and the second images
  indicates that the calculated time delay is 4.0 days earlier than
  {\it correct} value. Thus, the prediction of the third image is not
  very precise, but still feasible.}  Smaller error values lead to
better prediction.

Note that our assumptions of uncertainties are relatively
conservative, that is, the time delay error of actual observations
could be smaller than the current estimate. The uncertainty of time
delays for second and third images (5 days here), however, might be
longer. In order to check their dependences, we perform the same
analysis by changing the uncertainty of the time delay, for instance 3
days or 10 days, and find that the prediction changes only
$\mathcal{O}(0.01)$ day. Namely, the uncertainty of the time delay is
not essential for the prediction of the fourth image for this
particular case.

We also conduct the same analysis for other mock samples of
\cite{ogur10}. For this analysis, we use the following two criteria:
i) time delay between the first and the third images is longer than
five days; ii) time delay between the first and the fourth images is
longer than ten days. Then, we obtain 29 events from the mock
data. Among them, 23 events (79\%) present good agreements between the
estimated time delays and the correct values within one day.  Other
six events indicate small image separations ($\sim 0.1$ arcsec) or
short time delays between the first and the third images ($\sim 5$
days). By performing the analysis with smaller position and time delay
uncertainties than the previous one, i.e. 0.3 arcsec and 1 day, we
find that the prediction error is largely improved (within one day)
for five of them. One exceptional case (ID 69154345) has the
considerably delayed fourth image which appears 140.7 days after the
first image.  Therefore, after the detection of the third image, the
prediction of fourth image is doable in most cases.  Note that this
result may change with different assumption about the lens model, so
that a more systematic study is needed and will be presented in
following works.

For the current purpose, the fourth image needs to appear later than
10 days after the first image, in order to complete these analyses
before the fourth image emerges. In the mock catalog from
\cite{ogur10}, there are 50 Type Ib/c SNe, which have
$t_\mathrm{delay}$ (delay time between the first and fourth images)
$>$ 10 days and 24 SNe with $t_\mathrm{delay}>20$ days. Note that
their catalog is based on the rather stringent criterion that multiple
images be detected with more than 10$\sigma$ and be resolved by LSST
alone. By using other telescopes, which have better sensitivity and
angular resolution, the event rate can be increased by a factor of
10. Then, we may observe $\sim 1$ of these events per year in the LSST
era \citep[see also][]{gold17a}.

\section{Summary and discussion}
\label{sec:summary}

In the coming LSST era, we will observe $\mathcal{O}(100)$ of lensed
SNe. In this paper, we investigate the feasibility of predicting the
emergence of strongly lensed supernovae, based on mock data given by
\cite{ogur10}. Considering a family of lens models generating four
images, we found that it will be possible to predict the time and
position of the fourth SN image, given observations of the preceding
images.  In particular, if the separation of images is, roughly
speaking, larger than PSF (0.75 arcsec in this study) and the time
delay between the first and third image is longer than the observation
cadence (five days), the prediction of the forth image is possible
within one day.  The largest systematic error for evaluating the time
of SBO is LC modeling of the SN. We can reduce the error by conducting
detailed observations of the third image.

The microlensing effects would also introduce some error
\citep{dobl06}, which is not taken into account in this study.
Recently, \cite{gold17b} investigated the microlensing impact on LC of
Type Ia SNe, and showed that by making use of multi-band LCs the time
delay error by microlensing can be reduced to $\sim 1\%$ level.  A
similar study for Type Ib/c SNe is necessary to give uncertainties of
the prediction of SBO.

Not only SBO, also the early LC of SNe contains rich information. For
instance, with early X-ray emission, the explosion scenario of Type Ia
(single degenerate or double degenerate) can be distinguished
\citep{kase10b}. For core-collapse supernovae, the early LC provides
the very final stage of mass ejection from the progenitor stars, which
may change the features of SBO. If the SBO takes place in the dense
wind, the timescale would become longer \citep{tana16}. These features
can be tackled by multi-messenger observations of early LCs.

\section*{Acknowledgements}

We acknowledge M. Oguri for inspiring this work, providing their mock
data, and fruitful discussion. We also thank N. Tominaga for
discussing about optical transient observation and M. Werner for
insightful comments on the degrees of freedom of lens potential. This
study was supported in part by the Grant-in-Aid for Scientific
Research (Nos. 16H00869, 16K17665, and 17H02864), MEXT as ``Priority
Issue on Post-K computer'' (Elucidation of the Fundamental Laws and
Evolution of the Universe) and JICFuS.

\appendix
\section{glafic}
\label{sec:glafic}

For estimating delay time of the fourth image, we employ {\tt
  glafic}\footnote{http://www.slac.stanford.edu/\~{}oguri/glafic/}
version 1.2.8 with the following input file.

\begin{verbatim}
omega     0.260000
lambda	  0.740000
weos	  -1.000000
hubble	  0.720000
zl	  0.314000
prefix	  out
xmin	  -5.000000
ymin	  -5.000000
xmax	  5.000000
ymax	  5.000000
pix_ext   0.020000
pix_poi   0.500000
maxlev	  5

chi2_splane    0
chi2_checknimg 1
chi2_restart   -1
chi2_usemag    0

startup 2 0 1
lens sie    300.0  0.0 0.0 0.35   0.0 0.0 0.0
lens pert     2.0  0.0 0.0 0.05  60.0 0.0 0.0
point 0.87 0.0 0.0
end_startup

start_setopt
1 1 1 1 1 0 0 
1 0 0 1 1 0 0 
1 1 1
end_setopt

start_command

readobs_point obs_mock.dat
parprior prior_point.dat

optimize
findimg
\end{verbatim}

\end{document}